\documentclass[ amsmath,amssymb,aps,pra,reprint,floatfix]{revtex4-1}

\usepackage{graphicx}% Include figure files
\usepackage{dcolumn}% Align table columns on decimal point
\usepackage{bm}% bold math
\usepackage{bbm}
\usepackage{textcomp}
\usepackage{color}
\usepackage{float}
\begin{document}

\title{Procedure for systematically tuning up crosstalk in the cross resonance gate}
\author{Sarah Sheldon}
\author{Easwar Magesan}%
\author{Jerry M. Chow}
\author{Jay M. Gambetta}
\affiliation{IBM T.J. Watson Research Center, Yorktown Heights, NY 10598, USA}

\date{\today}% It is always \today, 

\newcommand{\ket}[1]{\left|#1\right>}
\newcommand{\scomment}[1]{\emph{\color[rgb]{0,0,0.6}(#1)}}

\begin{abstract}
We present improvements in both theoretical understanding and experimental implementation of the cross resonance (CR) gate that have led to shorter two-qubit gate times and interleaved randomized benchmarking fidelities exceeding 99\%.  The CR gate is an all-microwave two-qubit gate offers that does not require tunability and is therefore well suited to quantum computing architectures based on 2D superconducting qubits.  The performance of the gate has previously been hindered by long gate times and fidelities averaging 94-96\%.  We have developed a calibration procedure that accurately measures the full CR Hamiltonian.  The resulting measurements agree with theoretical analysis of the gate and also elucidate the error terms that have previously limited the gate fidelity.  The increase in fidelity that we have achieved was accomplished by introducing a second microwave drive tone on the target qubit to cancel unwanted components of the CR Hamiltonian.

\end{abstract}

\pacs{Valid PACS appear here}
\maketitle

The cross resonance (CR) gate is an entangling gate for superconducting qubits that uses only microwave control~\cite{rigetti,chow2011} and has been the standard for multi-qubit experiments in superconducting architectures using fixed-frequency transmon qubits~\cite{chow14,corcoles}.  Superconducting qubits arranged with shared quantum buses~\cite{majer} allow qubit networks to be designed with any desired connectivity. This flexiblity of design also translates into a flexibility of control and many choices in entangling gate implementations.  The CR gate is one choice of two-qubit gate that uses only microwave control, as opposed to using magnetic flux drives to tune two qubits into a specific resonance condition to entangle, as in the  controlled-Phase gate~\cite{barends,dicarlo}, or to tune a coupler directly~\cite{niskanen,bialczak,chen, mckay}. The CR gate requires a small static coupling of the qubit pair that slightly hybridizes the combined system and one additional microwave drive. 
The relatively low overhead of the CR scheme (the additional control line is combined with a single-qubit drive at room temperature) makes it an attractive gate for use in quantum computing architectures based on planar superconducting qubits.  Additionally, the CR gate is well-suited to transmon qubits~\cite{koch}, which have become the superconducting of choice due to promising long coherence and lifetimes~\cite{houck,barends2013}, limited charge noise~\cite{schreier}, and high single-qubit gate fidelities~\cite{sheldon}.  The microwave-only control allows the use of fixed-frequency transmons, further reducing the sources of possible noise~\cite{harlingen}.  Given all of these qualities, the CR gate has been a useful tool for application in multi-qubit experiments, including demonstrations of parity measurements required for the surface code~\cite{chow14}. 

Despite the appeal of the CR gate, its implementation has been hindered by slow gate times.  The CR gate relies on an always-on qubit-qubit coupling, but large couplings can lead to crosstalk between qubits.  This leads to a trade-off between fast, high-fidelity two-qubit gates and high-fidelity simultaneous single-qubit gates.  As a result, typical CR gates between transmon devices have resulted in gate times $>300\sim400$~ns, with measured fidelities of 94-96\%~\cite{corcolesRB,corcoles}.  
Here we describe improvements to the CR gate through a careful Hamiltonian analysis and novel tune-up procedure that reduce the gate time by a factor of two with corresponding fidelities  over 99\%. Our increased understanding of the CR Hamiltonian expands upon previous studies, which have primarily used a simple qubit model~\cite{paraoanu,rigetti}. We show that such models are incomplete and do not fully capture the dynamics of the complete two transmon qubit system.  

%\section{Randomized Benchmarking}
\begin{figure}[!ht]
	\centering
	\includegraphics[width = \columnwidth]{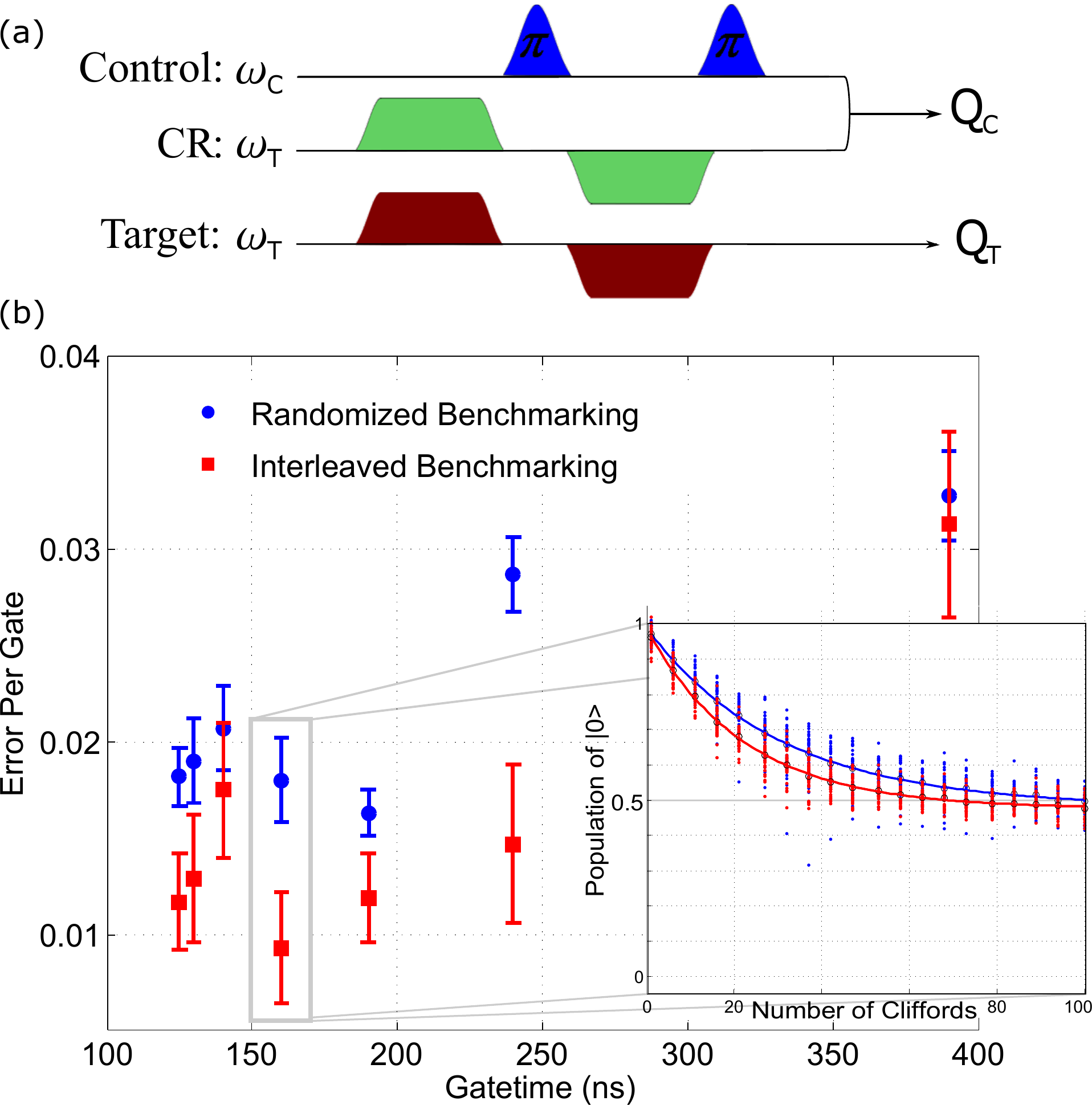}
	\caption{(Color online) (a) Schematic for the echoed cross resonance gate with active cancellation on the target qubit. The $\pi$-pulse on the control is a 20~ns  derivative removal via adiabatic
		gate (DRAG) pulse, and is buffered by two 10~ns delays.  The CR pulses are flat-topped Gaussian pulses, with 3$\sigma$ rise time where $\sigma = 5$~ns. (b)The error per gate found by randomized benchmarking (blue circles) and interleaved randomized benchmarking (red squares) as a function of gate time.  The benchmarking decay curves for highest fidelity point are highlighted in the inset. }
	\label{fig:CRbenchmarking}
\end{figure}

\begin{figure*}[!ht]
	\centering
	\includegraphics[width=2\columnwidth]{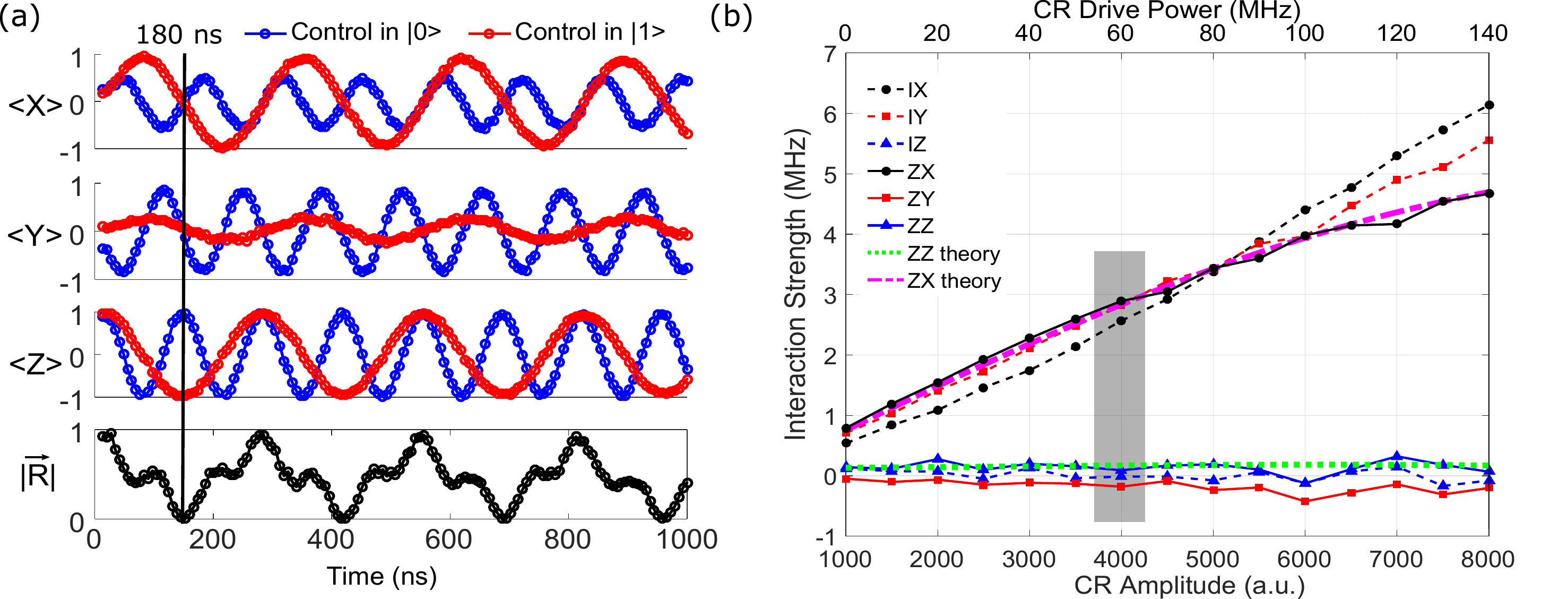}
	\caption{(Color online) (a) CR Rabi oscillations on the target qubit projected onto $x$, $y$, and $z$ for the control in $\ket{0}$ (blue) and the control in $\ket{1}$ (red).  The Bloch vector, $R$, for the target indicates gate times of maximal entanglement at points when $R = 0$.  (b) The CR Hamiltonian as a function of the two-qubit drive amplitude including IX (dashed black circles), IY (dashed red squares), IZ (dashed blue triangles), ZX (solid black circles), ZY (solid red squares), and ZZ (solid blue triangles). The magenta dot-dashed line and green dotted line correspond to the theory predictions for ZX and ZZ respectively as a function of the CR drive power on the upper x-axis.  The grey shaded region corresponds to the data from (a) and later in Fig. \ref{fig:CRphase}.}
	\label{fig:CR_HamiltonianVersusCRAmp}
\end{figure*}
The experimental system tested in this paper consists of two fixed-frequency transmon qubits coupled by a bus resonator.  The qubit frequencies are 4.914~GHz (target qubit) and 5.114~GHz (control qubit) with anharmonicities of -330~MHz for both, and the bus frequency is 6.31~GHz.  The coupling between the two qubits is estimated to be $J/2\pi = 3.8$~MHz for these parameters. The single-qubit gate fidelities measured with simultaneous randomized benchmarking~\cite{gambetta} are 0.9991$\pm$0.0002 for the target and 0.9992$\pm$0.0002 for control.  We characterize the two-qubit gate fidelities in the system by using interleaved randomized benchmarking (RB)~\cite{magesanIRB,gaebler}. For this measurement we first find the average fidelity per Clifford in the two-qubit system using standard randomized benchmarking~\cite{magesan12}, %\scomment{add emerson?}
and then repeat the measurement interleaving the CR gate between random Cliffords in the sequence.  The fidelities are extracted from exponential fits to the average over 35 random sequences each with a total length of 100 Clifford gates.  By applying the theoretical Hamiltonian understanding of the next few sections, we were able to benchmark results showing a CR gate fidelity, $f = 0.991\pm0.002$ for a 160~ns gate.  This gate time includes 20~ns added by the single qubit echo, which is buffered by two 10~ns delays. The CR pulses are rounded square pulses with Gaussian rise times of 15~ns.

The critical experimental technique for the improvement of the CR gate is an active cancellation pulse on the target qubit drive to eliminate unwanted interactions of the CR drive Hamiltonian, as shown in Fig.~\ref{fig:CRbenchmarking}(a). Without this cancellation tone the CR gate fidelity for the same gate time is $f = 0.948\pm 0.018$, demonstrating that the additional drive effectively cancels out the error terms of the CR Hamiltonian.

We measure the two-qubit gate fidelity as a function of gate time, plotted in Fig.~\ref{fig:CRbenchmarking}(b) with the RB curves for the gate with the best measured fidelity.  These gate times are the time for the total echoed CR gate including the single-qubit echoing gates. While the fidelities improve as the gate time decreases from 400~ns to 160~ns, shorter gates appear to saturate.  We suspect that leakage to higher levels or drive-induced dephasing from the strong two-qubit drive may be the source of this saturation.

The addition of the cancellation drive on the target is the consequence of measurements and simulation of the CR Hamiltonian indicating single-qubit errors that are not fully refocused by the standard CR gate (e.g. the echoed sequence in Fig.~\ref{fig:CRbenchmarking}[a] without the cancellation pulse on the target qubit). 
 We have developed an effective block-diagonal Hamiltonian model for the system dynamics under a CR drive that reveals the important contributions to the Hamiltonian on the qubit-qubit 4-dimensional subspace. The method finds a unitary operator $T$ that block-diagonalizes the full system Hamiltonian $H$, $H_\text{BD} :=T^\dagger HT$, under the constraint that $T$ minimizes the distance to the identity operation $\|T-\mathcal{I}\|_2$ (hence $T$ affects the system as little as possible). The solution for $T$ is given by
\begin{align}
T&= X\frac{X_\text{BD}}{\sqrt{X_\text{BD}X_\text{BD}^\dagger}},
\end{align}
where $X$ is the eigenvector matrix of H and is assumed to be non-singular, and $X_\text{BD}$ is the block-diagonal matrix of $X$.  
 
In our implementation the different blocks correspond to the different states of the control qubit and an off-resonant (higher energy) subspaces. The model predicts $ZX$ and $IX$ components of similar magnitude, negligible $IZ$ and $ZZ$ contributions, and a large $ZI$ term arising from a Stark shift of the control qubit from off-resonant driving. The complete CR Hamiltonian has the structure
\begin{equation}
\mathcal{H} = \frac{Z\otimes A}{2} + \frac{I\otimes B}{2}.
\label{eq:hamiltonian}
\end{equation}
 
Motivated by this understanding of the CR Hamiltonian, we have developed a protocol for experimentally measuring the CR Hamiltonian that allows us to determine the real error terms in the gate.  This measurement is accomplished by turning on a CR drive for some time and measuring the Rabi oscillations on the target qubit.  We project the target qubit state onto $x$, $y$, and $z$ following the Rabi drive and repeat for the control qubit in $\ket{0}$ and $\ket{1}$.  
%Because the Hamiltonian should only contain terms of the form $IU$ and $ZU$ for $U \in \{I,X,Y,Z\}$, this partial tomography on the control qubit is sufficient for determining the full two-qubit Hamiltonian and scales well as it only requires full tomography on the target qubit. 
 Fig.~\ref{fig:CR_HamiltonianVersusCRAmp}(a) contains an example of a complete set of such Hamiltonian tomography experiments.  The last parameter plotted, $\|\vec{R}\|$, is the Bloch vector of the target qubit defined as,
\begin{align}
\|\vec{R}\|& =\\ &\sqrt{\left(\left<X\right>_0+\left<X\right>_1\right)^2+\left(\left<Y\right>_0+\left<Y\right>_1\right)^2+\left(\left<Z\right>_0+\left<Z\right>_1\right)^2}\nonumber
\end{align} 
When this quantity goes to zero, the two-qubits are maximally entangled (unless the dynamics are completely mixed, but because $\|\vec{R}\|$ oscillates between zero and one we believe this is not the case).  We use $\|\vec{R}\|$ to estimate the gate length required to perform the entangling gate.
We fit the Rabi oscillations corresponding to the control in $\ket{0}$ and $\ket{1}$ separately with a Bloch equation model function:
\begin{equation}
\dot{\vec{r}}(t)= e^{At} \vec{r}(0),
\end{equation}
with the matrix $A$ defined as 
\begin{equation}
\left(\begin{array}{ccc}
0 & \Delta & \Omega_y\\
-\Delta & 0 & -\Omega_x\\
-\Omega_y & \Omega_x & 0
\end{array}\right).
\end{equation}
Here $\Delta$ is the control drive detuning, and $\Omega_{x,y}$ is the Rabi drive amplitude in along $\{x,y\}$. $\vec{r}(t)$ is the vector composed of the measured expectation values as a function of the length of the applied Rabi drive, $\left(\left<X(t)\right>,\left<Y(t)\right>,\left<Z(t)\right>\right)$. We find two generators corresponding to the control qubit in either $\ket{0}$ or $\ket{1}$, characterized by the vectors $$\vec{v}_{\{0,1\}} = (\Omega_x^{\{0,1\}},\Omega_y^{\{0,1\}},\Delta^{\{0,1\}}).$$  From these parameters we derive the CR drive Hamiltonian in terms of the six possible interactions: $IX$, $IY$, $IZ$, $ZX$, $ZY$, $ZZ$.  For example, $IX = (\Omega_x^0+\Omega_x^1)/2$ and $ZX = (\Omega_x^0-\Omega_x^1)/2$.
Note that this method of Hamiltonian tomography is applicable to any system with a Hamiltonian with the same form as Eq. \ref{eq:hamiltonian}. In addition, due to the symmetry of $\mathcal{H}$, this method scales efficiently for an $n$-qubit system since there are $n(n-1)/2$ different pairs and each pair requires six Rabi measurements described above.
\begin{figure}[!ht]
	\centering
	\includegraphics[width=0.9\columnwidth]{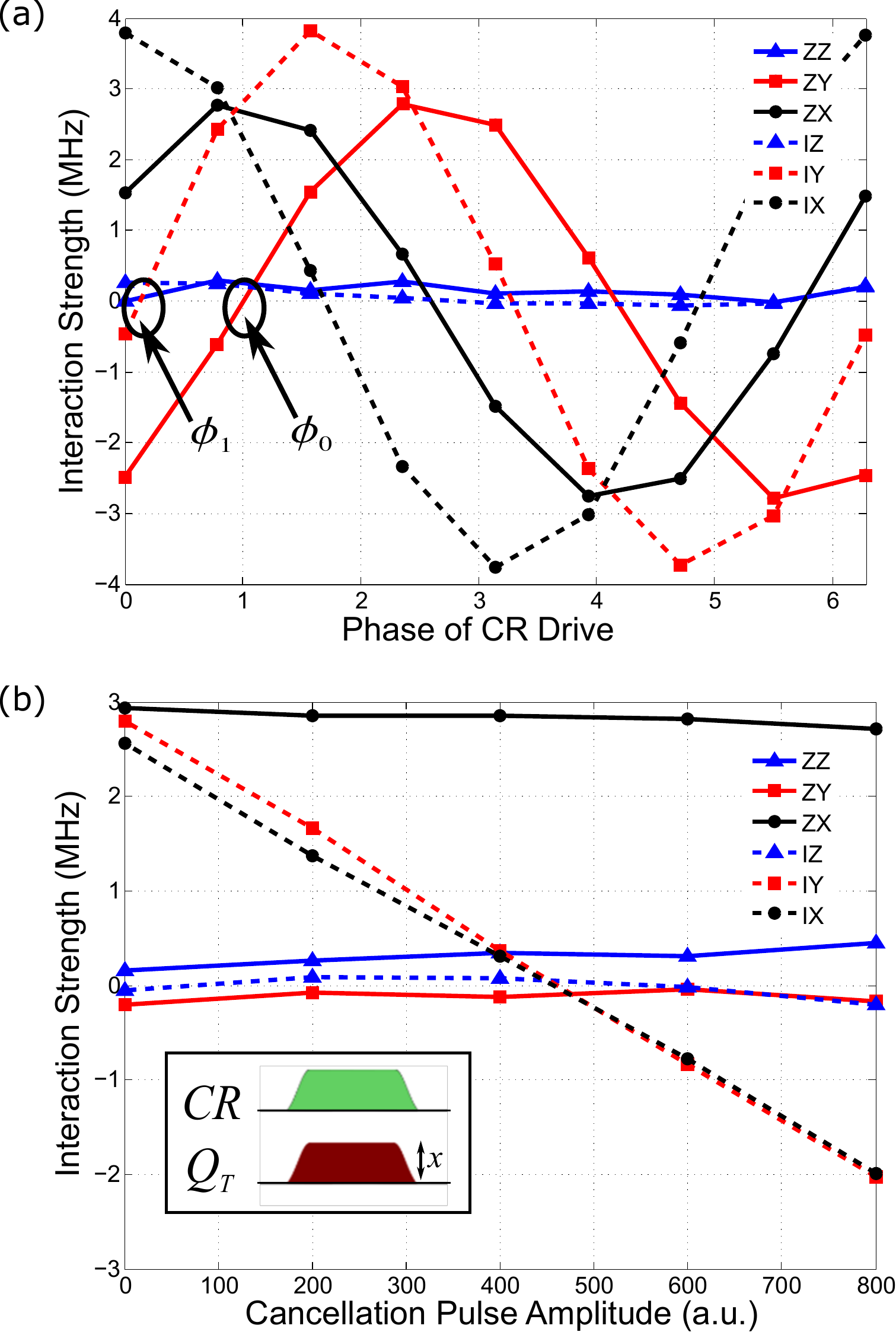}
	\caption{(Color online) (a) CR Hamiltonian parameters as a function of the drive phase: $ZZ$ (solid blue triangle), $ZY$ (solid red square), $ZX$ (solid black circle), $IZ$ (dashed blue triangle), $IY$ (dashed red square), $IX$ (dashed black circle).  (b) The same Hamiltonian parameters plotted versus the active cancellation pulse amplitude ($x$, as shown in the pulse schematic in the inset). The CR amplitude for this data corresponds to that of the shaded region in Fig. \ref{fig:CR_HamiltonianVersusCRAmp}(b)}
	\label{fig:CRphase}
\end{figure}

We can measure the CR Hamiltonian as a function of the two-qubit drive amplitude, as shown in Figure \ref{fig:CR_HamiltonianVersusCRAmp}.  
It is clear from this figure that components $IZ$ and $ZZ$ are small and independent of the drive power.  This data was acquired with the CR drive phase set so that the $ZY$ contribution is small and the conditional component of the Hamiltonian consists only of $ZX$.  The measured Hamiltonian parameters are consistent with the $ZX$ and $ZZ$ predicted by the effective Hamiltonian plotted in Fig.~\ref{fig:CR_HamiltonianVersusCRAmp}(b).

There is an unexpected feature in the experiment as there is also an $IY$ term present when the CR phase is set as above. We attribute this phase difference between conditional and single-qubit terms to classical crosstalk. Although such crosstalk has little effect on simultaneous RB since it is off-resonance, since the CR drive is applied at the frequency of the target qubit this crosstalk has a significant impact on the two-qubit gate calibration.  The standard CR gate is performed using an echo to refocus $IX$, $ZZ$, and $ZI$, depicted in Fig. \ref{fig:CRbenchmarking}(a).  The echoed scheme involves a $\pi$ pulse on the control qubit and a change of sign of the CR drive.  If $IY$ terms are present, however, then the echo fails to completely refocus the unwanted interactions. In this case the $IY$ interaction does not commute with $ZX$ and all higher-order terms of the commutator will be on during the two-qubit gate.

The gate calibration procedure is based on these Hamiltonian measurements. Ultimately the goal is to tune up a $ZX_{90}$, which is a generator of a controlled-NOT (CNOT) with single-qubit Clifford rotations.  The first step is to find the Hamiltonian parameters as a function of the phase of the CR drive (see Fig.~\ref{fig:CRphase}[a]).  We use Figure \ref{fig:CRphase} to find the CR phase, $\phi_0$ at which the $ZX$ component is maximized and $ZY$ is zero. 

With the CR phase fixed at $\phi_0$ the two-qubit drive produces nonzero $IX$ and $IY$ components.  We find the phase, $\phi_{1}$, at which the single-qubit component, $IY$, is zero.  The correct phase for the cancellation pulse is $\phi = \phi_0-\phi_1$, at which phase the single-qubit drive on the target matches $\tan^{-1}{\left(IY/IX\right)}$ for the two-qubit drive.

The second step is to set the cancellation pulse to the correct amplitude for canceling $\cos(\phi)IX +\sin(\phi)IY$.  Again we measure the CR Hamiltonian, this time sweeping the cancellation pulse amplitude as shown in Fig.~\ref{fig:CRphase}(b).  If $IX$ and $IY$ are not zero at the same cancellation amplitude ($x$ in the schematic shown in \ref{fig:CRphase}[b]), then the cancellation phase is incorrect.

%\section{Analysis of CR Gate Errors}
\begin{figure}[!htb]
	\centering
	\includegraphics[width = \columnwidth]{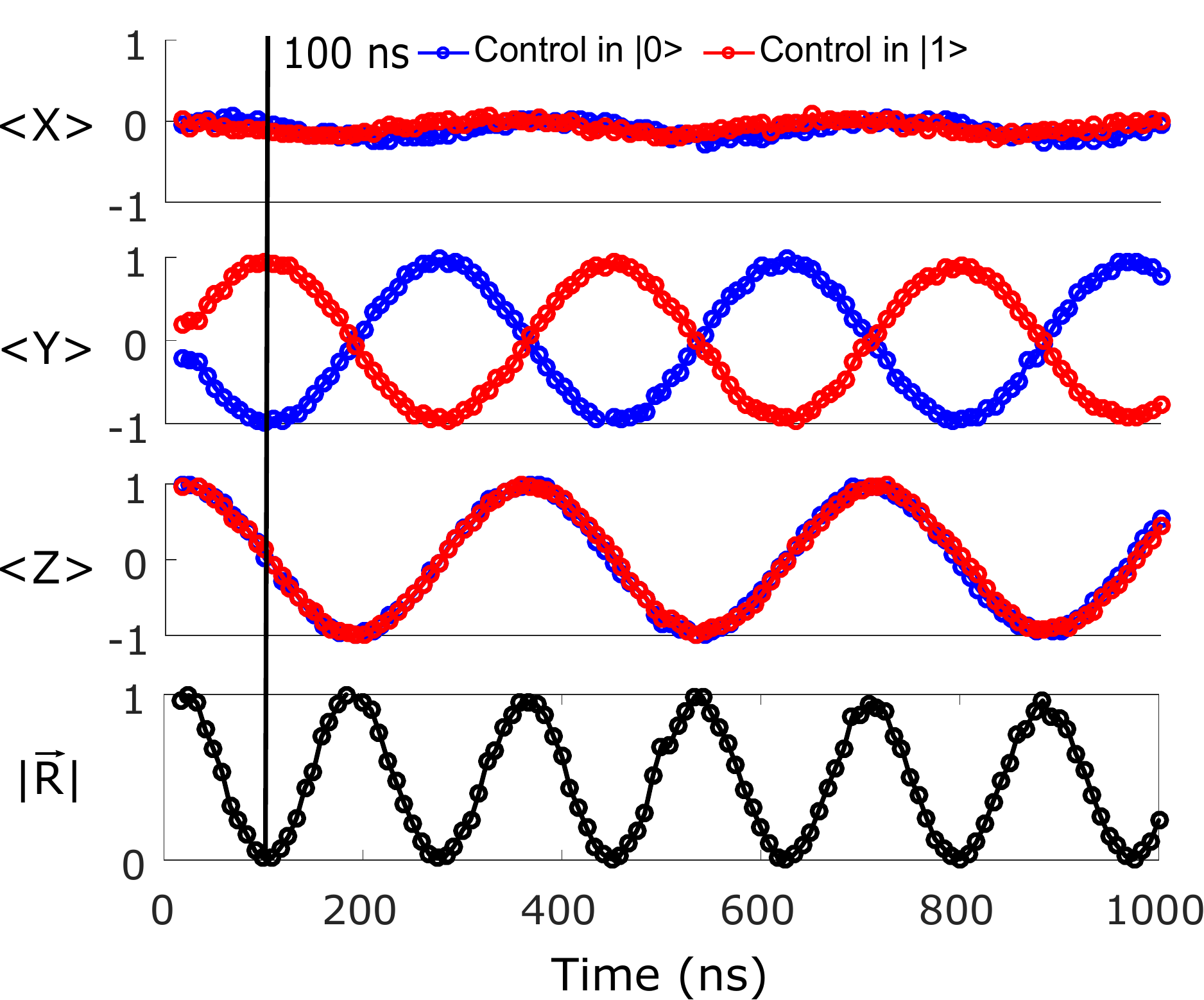}
	\caption{(Color online) CR Rabi experiments as a function of the CR pulsewidth after calibrating the active cancellation amplitude and phase.  From top to bottom are the expectation values $\left<X\right>$, $\left<Y\right>$,and $\left<Z\right>$, and the Bloch vector of the target qubit, $R$.}
	\label{fig:CR_Rabis_withCan}
\end{figure}
The CR-Rabi oscillations on the target qubit with the fully calibrated cancellation pulse (Fig. \ref{fig:CR_Rabis_withCan}) are much closer to the oscillations expected for a $ZX$ drive. From Fig. \ref{fig:CR_Rabis_withCan} we see that the entangling gate is significantly faster than previous cross resonance gates; the $R$-vector goes to zero at 100~ns for the particular CR amplitude used.  This is slightly slower than the entangling time of 83~ns that is predicted by the $ZX$ rate, which we attribute to the rise time of the pulse.  The echoed gate time of 160~ns is consistent with the R-vector measurement as it includes the additional single qubit gate and the rise and fall of both halves of the CR echo. Additionally the oscillation of $\|\vec{R}\|$ are sinusoidal, unlike in Fig.~\ref{fig:CR_HamiltonianVersusCRAmp}. If the cancellation were perfect, there would be no oscillation of $\left<X\right>$, and $\left<Y\right>$ oscillations would have full contrast.  
 \begin{figure}[!htb]
	\centering
	\includegraphics[width=\columnwidth]{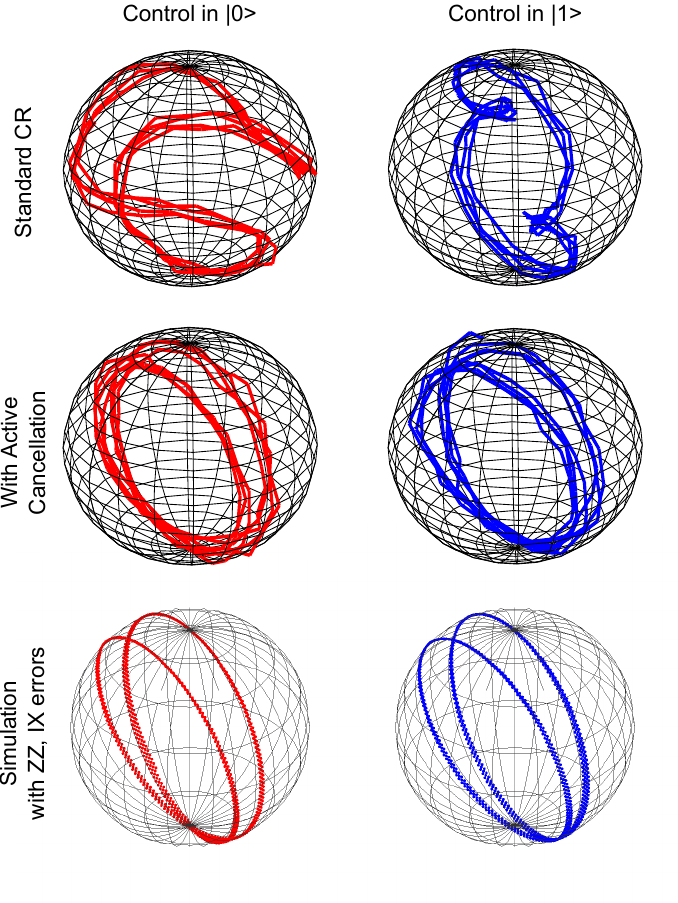}
	\caption{(Color online) Plots of the target qubit state on the Bloch sphere during evolution during the echoed cross resonance gate.  Left column is the target state when the control is in $\ket{0}$, with the right side corresponds to the control in $\ket{1}$.  Top row: echoed cross resonance gate with no active cancellation.  Middle: echoed cross resonance gate with active cancellation on the target qubit.  Bottom: simulations of the echoed cross resonance with small $IZ$ and $IX$ errors on the two-qubit drive. }
	\label{fig:blochsphere}
\end{figure}
Plotting the same data on the Bloch sphere, as in Fig. \ref{fig:blochsphere}, we can see the trajectory of the target qubit during the CR gate.  The data in the top row of Fig. \ref{fig:blochsphere} is the target trajectory when the two-qubit echoed gate is applied with no active cancellation pulse.  A perfect $ZX$ gate would result in a circle on the surface of the Bloch sphere, but the echoed CR gate creates a more convoluted path on the Bloch sphere before an entangling gate is achieved.  When the cancellation pulse is added (middle row), the picture is qualitatively closer to the ideal circle on the Bloch sphere, but still some error remains.  Comparing to a simulated echoed gate (bottom row), we find that the residual error is consistent with a $ZZ$ on the order of the $ZZ$ error we measure in the calibration sweeps and an order of magnitude smaller $IX$.

While the measured fidelity is higher than previously reported fidelities for CR two-qubit gates, it is still not yet limited by coherence.  The limit placed on the fidelity by $T_1$ (38$\pm$2~$\mu$s/41$\pm$2~$\mu$s for the control/target) and $T_2$ (50$\pm$4~$\mu$s/61$\pm$6~$\mu$s) for the two-qubits is 0.996.  There is evidence from studies of single-qubit gates that drive-activated dephasing may be a limiting factor for the error rates of gates at similar drive powers~\cite{sheldon,ball}.  Even so, it appears from the CR Rabi data that coherent errors have not been fully eliminated in the CR gate.

While it is clear that some coherent errors remain on the cross resonance gate, the inclusion of an active cancellation tone has produced a dramatic improvement in the two-qubit gate fidelity.  The calibration procedure that we have developed has provided insight into the full CR Hamiltonian, and revealed that a single-qubit phase shift due to crosstalk (either classical or quantum) is a significant source of error in the echoed CR gate.  Future work will focus on shortening the gate times further by tuning up a CR gate without an echo and developing faster and more robust calibration procedures.

These improvements to the microwave-driven CR gate are encouraging for a quantum computing architecture built out of fixed-frequency qubits.  With greater knowledge of the CR Hamiltonian, we are confident that further improvements can be made to reduce the two-qubit gate time and increase fidelity.

We acknowledge contributions from Marcus Brink, Jim Rozen, and Jack Rohrs. This work was supported by ARO under contract W911NF-14-1-0124.

\bibliography{crossResonance}

\begin{thebibliography}{24}%
\makeatletter
\providecommand \@ifxundefined [1]{%
 \@ifx{#1\undefined}
}%
\providecommand \@ifnum [1]{%
 \ifnum #1\expandafter \@firstoftwo
 \else \expandafter \@secondoftwo
 \fi
}%
\providecommand \@ifx [1]{%
 \ifx #1\expandafter \@firstoftwo
 \else \expandafter \@secondoftwo
 \fi
}%
\providecommand \natexlab [1]{#1}%
\providecommand \enquote  [1]{``#1''}%
\providecommand \bibnamefont  [1]{#1}%
\providecommand \bibfnamefont [1]{#1}%
\providecommand \citenamefont [1]{#1}%
\providecommand \href@noop [0]{\@secondoftwo}%
\providecommand \href [0]{\begingroup \@sanitize@url \@href}%
\providecommand \@href[1]{\@@startlink{#1}\@@href}%
\providecommand \@@href[1]{\endgroup#1\@@endlink}%
\providecommand \@sanitize@url [0]{\catcode `\\12\catcode `\$12\catcode
  `\&12\catcode `\#12\catcode `\^12\catcode `\_12\catcode `\%12\relax}%
\providecommand \@@startlink[1]{}%
\providecommand \@@endlink[0]{}%
\providecommand \url  [0]{\begingroup\@sanitize@url \@url }%
\providecommand \@url [1]{\endgroup\@href {#1}{\urlprefix }}%
\providecommand \urlprefix  [0]{URL }%
\providecommand \Eprint [0]{\href }%
\providecommand \doibase [0]{http://dx.doi.org/}%
\providecommand \selectlanguage [0]{\@gobble}%
\providecommand \bibinfo  [0]{\@secondoftwo}%
\providecommand \bibfield  [0]{\@secondoftwo}%
\providecommand \translation [1]{[#1]}%
\providecommand \BibitemOpen [0]{}%
\providecommand \bibitemStop [0]{}%
\providecommand \bibitemNoStop [0]{.\EOS\space}%
\providecommand \EOS [0]{\spacefactor3000\relax}%
\providecommand \BibitemShut  [1]{\csname bibitem#1\endcsname}%
\let\auto@bib@innerbib\@empty
%</preamble>
\bibitem [{\citenamefont {Rigetti}\ and\ \citenamefont
  {Devoret}(2010)}]{rigetti}%
  \BibitemOpen
  \bibfield  {author} {\bibinfo {author} {\bibfnamefont {C.}~\bibnamefont
  {Rigetti}}\ and\ \bibinfo {author} {\bibfnamefont {M.}~\bibnamefont
  {Devoret}},\ }\href {\doibase 10.1103/PhysRevB.81.134507} {\bibfield
  {journal} {\bibinfo  {journal} {Phys. Rev. B}\ }\textbf {\bibinfo {volume}
  {81}},\ \bibinfo {pages} {134507} (\bibinfo {year} {2010})}\BibitemShut
  {NoStop}%
\bibitem [{\citenamefont {Chow}\ \emph {et~al.}(2011)\citenamefont {Chow},
  \citenamefont {C\'orcoles}, \citenamefont {Gambetta}, \citenamefont
  {Rigetti}, \citenamefont {Johnson}, \citenamefont {Smolin}, \citenamefont
  {Rozen}, \citenamefont {Keefe}, \citenamefont {Rothwell}, \citenamefont
  {Ketchen},\ and\ \citenamefont {Steffen}}]{chow2011}%
  \BibitemOpen
  \bibfield  {author} {\bibinfo {author} {\bibfnamefont {J.~M.}\ \bibnamefont
  {Chow}}, \bibinfo {author} {\bibfnamefont {A.~D.}\ \bibnamefont
  {C\'orcoles}}, \bibinfo {author} {\bibfnamefont {J.~M.}\ \bibnamefont
  {Gambetta}}, \bibinfo {author} {\bibfnamefont {C.}~\bibnamefont {Rigetti}},
  \bibinfo {author} {\bibfnamefont {B.~R.}\ \bibnamefont {Johnson}}, \bibinfo
  {author} {\bibfnamefont {J.~A.}\ \bibnamefont {Smolin}}, \bibinfo {author}
  {\bibfnamefont {J.~R.}\ \bibnamefont {Rozen}}, \bibinfo {author}
  {\bibfnamefont {G.~A.}\ \bibnamefont {Keefe}}, \bibinfo {author}
  {\bibfnamefont {M.~B.}\ \bibnamefont {Rothwell}}, \bibinfo {author}
  {\bibfnamefont {M.~B.}\ \bibnamefont {Ketchen}}, \ and\ \bibinfo {author}
  {\bibfnamefont {M.}~\bibnamefont {Steffen}},\ }\href {\doibase
  10.1103/PhysRevLett.107.080502} {\bibfield  {journal} {\bibinfo  {journal}
  {Phys. Rev. Lett.}\ }\textbf {\bibinfo {volume} {107}},\ \bibinfo {pages}
  {080502} (\bibinfo {year} {2011})}\BibitemShut {NoStop}%
\bibitem [{\citenamefont {{Chow}}\ \emph {et~al.}(2014)\citenamefont {{Chow}},
  \citenamefont {{Gambetta}}, \citenamefont {{Magesan}}, \citenamefont
  {{Abraham}}, \citenamefont {{Cross}}, \citenamefont {{Johnson}},
  \citenamefont {{Masluk}}, \citenamefont {{Ryan}}, \citenamefont {{Smolin}},
  \citenamefont {{Srinivasan}},\ and\ \citenamefont {{Steffen}}}]{chow14}%
  \BibitemOpen
  \bibfield  {author} {\bibinfo {author} {\bibfnamefont {J.~M.}\ \bibnamefont
  {{Chow}}}, \bibinfo {author} {\bibfnamefont {J.~M.}\ \bibnamefont
  {{Gambetta}}}, \bibinfo {author} {\bibfnamefont {E.}~\bibnamefont
  {{Magesan}}}, \bibinfo {author} {\bibfnamefont {D.~W.}\ \bibnamefont
  {{Abraham}}}, \bibinfo {author} {\bibfnamefont {A.~W.}\ \bibnamefont
  {{Cross}}}, \bibinfo {author} {\bibfnamefont {B.~R.}\ \bibnamefont
  {{Johnson}}}, \bibinfo {author} {\bibfnamefont {N.~A.}\ \bibnamefont
  {{Masluk}}}, \bibinfo {author} {\bibfnamefont {C.~A.}\ \bibnamefont
  {{Ryan}}}, \bibinfo {author} {\bibfnamefont {J.~A.}\ \bibnamefont
  {{Smolin}}}, \bibinfo {author} {\bibfnamefont {S.~J.}\ \bibnamefont
  {{Srinivasan}}}, \ and\ \bibinfo {author} {\bibfnamefont {M.}~\bibnamefont
  {{Steffen}}},\ }\href {\doibase 10.1038/ncomms5015} {\bibfield  {journal}
  {\bibinfo  {journal} {Nature Communications}\ }\textbf {\bibinfo {volume}
  {5}},\ \bibinfo {eid} {4015} (\bibinfo {year} {2014})}\BibitemShut {NoStop}%
\bibitem [{\citenamefont {Corcoles}\ \emph {et~al.}(2015)\citenamefont
  {Corcoles}, \citenamefont {Magesan}, \citenamefont {Srinivasan},
  \citenamefont {Cross}, \citenamefont {Steffen}, \citenamefont {Gambetta},\
  and\ \citenamefont {Chow}}]{corcoles}%
  \BibitemOpen
  \bibfield  {author} {\bibinfo {author} {\bibfnamefont {A.}~\bibnamefont
  {Corcoles}}, \bibinfo {author} {\bibfnamefont {E.}~\bibnamefont {Magesan}},
  \bibinfo {author} {\bibfnamefont {S.~J.}\ \bibnamefont {Srinivasan}},
  \bibinfo {author} {\bibfnamefont {A.~W.}\ \bibnamefont {Cross}}, \bibinfo
  {author} {\bibfnamefont {M.}~\bibnamefont {Steffen}}, \bibinfo {author}
  {\bibfnamefont {J.~M.}\ \bibnamefont {Gambetta}}, \ and\ \bibinfo {author}
  {\bibfnamefont {J.~M.}\ \bibnamefont {Chow}},\ }\href {\doibase
  10.1038/ncomms7979} {\bibfield  {journal} {\bibinfo  {journal} {Nature
  Communications}\ }\textbf {\bibinfo {volume} {6}} (\bibinfo {year} {2015}),\
  10.1038/ncomms7979}\BibitemShut {NoStop}%
\bibitem [{\citenamefont {Majer}\ \emph {et~al.}(2007)\citenamefont {Majer},
  \citenamefont {Chow}, \citenamefont {Gambetta}, \citenamefont {Koch},
  \citenamefont {Johnson}, \citenamefont {Schreier}, \citenamefont {Frunzio},
  \citenamefont {Schuster}, \citenamefont {Houck}, \citenamefont {Wallraff},
  \citenamefont {Blais}, \citenamefont {Devoret}, \citenamefont {Girvin},\ and\
  \citenamefont {Schoelkopf}}]{majer}%
  \BibitemOpen
  \bibfield  {author} {\bibinfo {author} {\bibfnamefont {J.}~\bibnamefont
  {Majer}}, \bibinfo {author} {\bibfnamefont {J.~M.}\ \bibnamefont {Chow}},
  \bibinfo {author} {\bibfnamefont {J.~M.}\ \bibnamefont {Gambetta}}, \bibinfo
  {author} {\bibfnamefont {J.}~\bibnamefont {Koch}}, \bibinfo {author}
  {\bibfnamefont {B.~R.}\ \bibnamefont {Johnson}}, \bibinfo {author}
  {\bibfnamefont {J.~A.}\ \bibnamefont {Schreier}}, \bibinfo {author}
  {\bibfnamefont {L.}~\bibnamefont {Frunzio}}, \bibinfo {author} {\bibfnamefont
  {D.~I.}\ \bibnamefont {Schuster}}, \bibinfo {author} {\bibfnamefont {A.~A.}\
  \bibnamefont {Houck}}, \bibinfo {author} {\bibfnamefont {A.}~\bibnamefont
  {Wallraff}}, \bibinfo {author} {\bibfnamefont {A.}~\bibnamefont {Blais}},
  \bibinfo {author} {\bibfnamefont {M.~H.}\ \bibnamefont {Devoret}}, \bibinfo
  {author} {\bibfnamefont {S.~M.}\ \bibnamefont {Girvin}}, \ and\ \bibinfo
  {author} {\bibfnamefont {R.~J.}\ \bibnamefont {Schoelkopf}},\ }\href@noop {}
  {\bibfield  {journal} {\bibinfo  {journal} {Nature}\ }\textbf {\bibinfo
  {volume} {449}},\ \bibinfo {pages} {443} (\bibinfo {year}
  {2007})}\BibitemShut {NoStop}%
\bibitem [{\citenamefont {Barends}\ \emph {et~al.}(2014)\citenamefont
  {Barends}, \citenamefont {Kelly}, \citenamefont {Megrant}, \citenamefont
  {Veitia}, \citenamefont {Sank}, \citenamefont {Jeffrey}, \citenamefont
  {White}, \citenamefont {Mutus}, \citenamefont {Fowler}, \citenamefont
  {Campbell}, \citenamefont {Chen}, \citenamefont {Chen}, \citenamefont
  {Chiaro}, \citenamefont {Dunsworth}, \citenamefont {Neill}, \citenamefont
  {O/'Malley}, \citenamefont {Roushan}, \citenamefont {Vainsencher},
  \citenamefont {Wenner}, \citenamefont {Korotkov}, \citenamefont {Cleland},\
  and\ \citenamefont {Martinis}}]{barends}%
  \BibitemOpen
  \bibfield  {author} {\bibinfo {author} {\bibfnamefont {R.}~\bibnamefont
  {Barends}}, \bibinfo {author} {\bibfnamefont {J.}~\bibnamefont {Kelly}},
  \bibinfo {author} {\bibfnamefont {A.}~\bibnamefont {Megrant}}, \bibinfo
  {author} {\bibfnamefont {A.}~\bibnamefont {Veitia}}, \bibinfo {author}
  {\bibfnamefont {D.}~\bibnamefont {Sank}}, \bibinfo {author} {\bibfnamefont
  {E.}~\bibnamefont {Jeffrey}}, \bibinfo {author} {\bibfnamefont {T.~C.}\
  \bibnamefont {White}}, \bibinfo {author} {\bibfnamefont {J.}~\bibnamefont
  {Mutus}}, \bibinfo {author} {\bibfnamefont {A.~G.}\ \bibnamefont {Fowler}},
  \bibinfo {author} {\bibfnamefont {B.}~\bibnamefont {Campbell}}, \bibinfo
  {author} {\bibfnamefont {Y.}~\bibnamefont {Chen}}, \bibinfo {author}
  {\bibfnamefont {Z.}~\bibnamefont {Chen}}, \bibinfo {author} {\bibfnamefont
  {B.}~\bibnamefont {Chiaro}}, \bibinfo {author} {\bibfnamefont
  {A.}~\bibnamefont {Dunsworth}}, \bibinfo {author} {\bibfnamefont
  {C.}~\bibnamefont {Neill}}, \bibinfo {author} {\bibfnamefont
  {P.}~\bibnamefont {O/'Malley}}, \bibinfo {author} {\bibfnamefont
  {P.}~\bibnamefont {Roushan}}, \bibinfo {author} {\bibfnamefont
  {A.}~\bibnamefont {Vainsencher}}, \bibinfo {author} {\bibfnamefont
  {J.}~\bibnamefont {Wenner}}, \bibinfo {author} {\bibfnamefont {A.~N.}\
  \bibnamefont {Korotkov}}, \bibinfo {author} {\bibfnamefont {A.~N.}\
  \bibnamefont {Cleland}}, \ and\ \bibinfo {author} {\bibfnamefont {J.~M.}\
  \bibnamefont {Martinis}},\ }\href {\doibase 10.1038/nature13171} {\bibfield
  {journal} {\bibinfo  {journal} {Nature}\ }\textbf {\bibinfo {volume} {508}},\
  \bibinfo {pages} {500} (\bibinfo {year} {2014})}\BibitemShut {NoStop}%
\bibitem [{\citenamefont {DiCarlo}\ \emph {et~al.}(2009)\citenamefont
  {DiCarlo}, \citenamefont {Chow}, \citenamefont {Gambetta}, \citenamefont
  {Bishop}, \citenamefont {Johnson}, \citenamefont {Schuster}, \citenamefont
  {Majer}, \citenamefont {Blais}, \citenamefont {Frunzio}, \citenamefont
  {Girvin},\ and\ \citenamefont {Schoelkopf}}]{dicarlo}%
  \BibitemOpen
  \bibfield  {author} {\bibinfo {author} {\bibfnamefont {L.}~\bibnamefont
  {DiCarlo}}, \bibinfo {author} {\bibfnamefont {J.~M.}\ \bibnamefont {Chow}},
  \bibinfo {author} {\bibfnamefont {J.~M.}\ \bibnamefont {Gambetta}}, \bibinfo
  {author} {\bibfnamefont {L.~S.}\ \bibnamefont {Bishop}}, \bibinfo {author}
  {\bibfnamefont {B.~R.}\ \bibnamefont {Johnson}}, \bibinfo {author}
  {\bibfnamefont {D.~I.}\ \bibnamefont {Schuster}}, \bibinfo {author}
  {\bibfnamefont {J.}~\bibnamefont {Majer}}, \bibinfo {author} {\bibfnamefont
  {A.}~\bibnamefont {Blais}}, \bibinfo {author} {\bibfnamefont
  {L.}~\bibnamefont {Frunzio}}, \bibinfo {author} {\bibfnamefont {S.~M.}\
  \bibnamefont {Girvin}}, \ and\ \bibinfo {author} {\bibfnamefont {R.~J.}\
  \bibnamefont {Schoelkopf}},\ }\href@noop {} {\bibfield  {journal} {\bibinfo
  {journal} {Nature}\ }\textbf {\bibinfo {volume} {460}},\ \bibinfo {pages}
  {240} (\bibinfo {year} {2009})}\BibitemShut {NoStop}%
\bibitem [{\citenamefont {Niskanen}\ \emph {et~al.}(2007)\citenamefont
  {Niskanen}, \citenamefont {Harrabi}, \citenamefont {Yoshihara}, \citenamefont
  {Nakamura}, \citenamefont {Lloyd},\ and\ \citenamefont {Tsai}}]{niskanen}%
  \BibitemOpen
  \bibfield  {author} {\bibinfo {author} {\bibfnamefont {A.~O.}\ \bibnamefont
  {Niskanen}}, \bibinfo {author} {\bibfnamefont {K.}~\bibnamefont {Harrabi}},
  \bibinfo {author} {\bibfnamefont {F.}~\bibnamefont {Yoshihara}}, \bibinfo
  {author} {\bibfnamefont {Y.}~\bibnamefont {Nakamura}}, \bibinfo {author}
  {\bibfnamefont {S.}~\bibnamefont {Lloyd}}, \ and\ \bibinfo {author}
  {\bibfnamefont {J.~S.}\ \bibnamefont {Tsai}},\ }\href {\doibase
  10.1126/science.1141324} {\bibfield  {journal} {\bibinfo  {journal}
  {Science}\ }\textbf {\bibinfo {volume} {316}},\ \bibinfo {pages} {723}
  (\bibinfo {year} {2007})}\BibitemShut {NoStop}%
\bibitem [{\citenamefont {Bialczak}\ \emph {et~al.}(2011)\citenamefont
  {Bialczak}, \citenamefont {Ansmann}, \citenamefont {Hofheinz}, \citenamefont
  {Lenander}, \citenamefont {Lucero}, \citenamefont {Neeley}, \citenamefont
  {O'Connell}, \citenamefont {Sank}, \citenamefont {Wang}, \citenamefont
  {Weides}, \citenamefont {Wenner}, \citenamefont {Yamamoto}, \citenamefont
  {Cleland},\ and\ \citenamefont {Martinis}}]{bialczak}%
  \BibitemOpen
  \bibfield  {author} {\bibinfo {author} {\bibfnamefont {R.~C.}\ \bibnamefont
  {Bialczak}}, \bibinfo {author} {\bibfnamefont {M.}~\bibnamefont {Ansmann}},
  \bibinfo {author} {\bibfnamefont {M.}~\bibnamefont {Hofheinz}}, \bibinfo
  {author} {\bibfnamefont {M.}~\bibnamefont {Lenander}}, \bibinfo {author}
  {\bibfnamefont {E.}~\bibnamefont {Lucero}}, \bibinfo {author} {\bibfnamefont
  {M.}~\bibnamefont {Neeley}}, \bibinfo {author} {\bibfnamefont {A.~D.}\
  \bibnamefont {O'Connell}}, \bibinfo {author} {\bibfnamefont {D.}~\bibnamefont
  {Sank}}, \bibinfo {author} {\bibfnamefont {H.}~\bibnamefont {Wang}}, \bibinfo
  {author} {\bibfnamefont {M.}~\bibnamefont {Weides}}, \bibinfo {author}
  {\bibfnamefont {J.}~\bibnamefont {Wenner}}, \bibinfo {author} {\bibfnamefont
  {T.}~\bibnamefont {Yamamoto}}, \bibinfo {author} {\bibfnamefont {A.~N.}\
  \bibnamefont {Cleland}}, \ and\ \bibinfo {author} {\bibfnamefont {J.~M.}\
  \bibnamefont {Martinis}},\ }\href {\doibase 10.1103/PhysRevLett.106.060501}
  {\bibfield  {journal} {\bibinfo  {journal} {Phys. Rev. Lett.}\ }\textbf
  {\bibinfo {volume} {106}},\ \bibinfo {pages} {060501} (\bibinfo {year}
  {2011})}\BibitemShut {NoStop}%
\bibitem [{\citenamefont {Chen}\ \emph {et~al.}(2014)\citenamefont {Chen},
  \citenamefont {Neill}, \citenamefont {Roushan}, \citenamefont {Leung},
  \citenamefont {Fang}, \citenamefont {Barends}, \citenamefont {Kelly},
  \citenamefont {Campbell}, \citenamefont {Chen}, \citenamefont {Chiaro},
  \citenamefont {Dunsworth}, \citenamefont {Jeffrey}, \citenamefont {Megrant},
  \citenamefont {Mutus}, \citenamefont {O'Malley}, \citenamefont {Quintana},
  \citenamefont {Sank}, \citenamefont {Vainsencher}, \citenamefont {Wenner},
  \citenamefont {White}, \citenamefont {Geller}, \citenamefont {Cleland},\ and\
  \citenamefont {Martinis}}]{chen}%
  \BibitemOpen
  \bibfield  {author} {\bibinfo {author} {\bibfnamefont {Y.}~\bibnamefont
  {Chen}}, \bibinfo {author} {\bibfnamefont {C.}~\bibnamefont {Neill}},
  \bibinfo {author} {\bibfnamefont {P.}~\bibnamefont {Roushan}}, \bibinfo
  {author} {\bibfnamefont {N.}~\bibnamefont {Leung}}, \bibinfo {author}
  {\bibfnamefont {M.}~\bibnamefont {Fang}}, \bibinfo {author} {\bibfnamefont
  {R.}~\bibnamefont {Barends}}, \bibinfo {author} {\bibfnamefont
  {J.}~\bibnamefont {Kelly}}, \bibinfo {author} {\bibfnamefont
  {B.}~\bibnamefont {Campbell}}, \bibinfo {author} {\bibfnamefont
  {Z.}~\bibnamefont {Chen}}, \bibinfo {author} {\bibfnamefont {B.}~\bibnamefont
  {Chiaro}}, \bibinfo {author} {\bibfnamefont {A.}~\bibnamefont {Dunsworth}},
  \bibinfo {author} {\bibfnamefont {E.}~\bibnamefont {Jeffrey}}, \bibinfo
  {author} {\bibfnamefont {A.}~\bibnamefont {Megrant}}, \bibinfo {author}
  {\bibfnamefont {J.~Y.}\ \bibnamefont {Mutus}}, \bibinfo {author}
  {\bibfnamefont {P.~J.~J.}\ \bibnamefont {O'Malley}}, \bibinfo {author}
  {\bibfnamefont {C.~M.}\ \bibnamefont {Quintana}}, \bibinfo {author}
  {\bibfnamefont {D.}~\bibnamefont {Sank}}, \bibinfo {author} {\bibfnamefont
  {A.}~\bibnamefont {Vainsencher}}, \bibinfo {author} {\bibfnamefont
  {J.}~\bibnamefont {Wenner}}, \bibinfo {author} {\bibfnamefont {T.~C.}\
  \bibnamefont {White}}, \bibinfo {author} {\bibfnamefont {M.~R.}\ \bibnamefont
  {Geller}}, \bibinfo {author} {\bibfnamefont {A.~N.}\ \bibnamefont {Cleland}},
  \ and\ \bibinfo {author} {\bibfnamefont {J.~M.}\ \bibnamefont {Martinis}},\
  }\href {\doibase 10.1103/PhysRevLett.113.220502} {\bibfield  {journal}
  {\bibinfo  {journal} {Phys. Rev. Lett.}\ }\textbf {\bibinfo {volume} {113}},\
  \bibinfo {pages} {220502} (\bibinfo {year} {2014})}\BibitemShut {NoStop}%
\bibitem [{\citenamefont {McKay}\ \emph {et~al.}(tion)\citenamefont {McKay}
  \emph {et~al.}}]{mckay}%
  \BibitemOpen
  \bibfield  {author} {\bibinfo {author} {\bibfnamefont {D.~C.}\ \bibnamefont
  {McKay}} \emph {et~al.},\ }\href@noop {} {\  (\bibinfo {year} {in
  preparation})}\BibitemShut {NoStop}%
\bibitem [{\citenamefont {Koch}\ \emph {et~al.}(2007)\citenamefont {Koch},
  \citenamefont {Yu}, \citenamefont {Gambetta}, \citenamefont {Houck},
  \citenamefont {Schuster}, \citenamefont {Majer}, \citenamefont {Blais},
  \citenamefont {Devoret}, \citenamefont {Girvin},\ and\ \citenamefont
  {Schoelkopf}}]{koch}%
  \BibitemOpen
  \bibfield  {author} {\bibinfo {author} {\bibfnamefont {J.}~\bibnamefont
  {Koch}}, \bibinfo {author} {\bibfnamefont {T.~M.}\ \bibnamefont {Yu}},
  \bibinfo {author} {\bibfnamefont {J.}~\bibnamefont {Gambetta}}, \bibinfo
  {author} {\bibfnamefont {A.~A.}\ \bibnamefont {Houck}}, \bibinfo {author}
  {\bibfnamefont {D.~I.}\ \bibnamefont {Schuster}}, \bibinfo {author}
  {\bibfnamefont {J.}~\bibnamefont {Majer}}, \bibinfo {author} {\bibfnamefont
  {A.}~\bibnamefont {Blais}}, \bibinfo {author} {\bibfnamefont {M.~H.}\
  \bibnamefont {Devoret}}, \bibinfo {author} {\bibfnamefont {S.~M.}\
  \bibnamefont {Girvin}}, \ and\ \bibinfo {author} {\bibfnamefont {R.~J.}\
  \bibnamefont {Schoelkopf}},\ }\href {\doibase 10.1103/PhysRevA.76.042319}
  {\bibfield  {journal} {\bibinfo  {journal} {Phys. Rev. A}\ }\textbf {\bibinfo
  {volume} {76}},\ \bibinfo {pages} {042319} (\bibinfo {year}
  {2007})}\BibitemShut {NoStop}%
\bibitem [{\citenamefont {Houck}\ \emph {et~al.}(2008)\citenamefont {Houck},
  \citenamefont {Schreier}, \citenamefont {Johnson}, \citenamefont {Chow},
  \citenamefont {Koch}, \citenamefont {Gambetta}, \citenamefont {Schuster},
  \citenamefont {Frunzio}, \citenamefont {Devoret}, \citenamefont {Girvin},\
  and\ \citenamefont {Schoelkopf}}]{houck}%
  \BibitemOpen
  \bibfield  {author} {\bibinfo {author} {\bibfnamefont {A.~A.}\ \bibnamefont
  {Houck}}, \bibinfo {author} {\bibfnamefont {J.~A.}\ \bibnamefont {Schreier}},
  \bibinfo {author} {\bibfnamefont {B.~R.}\ \bibnamefont {Johnson}}, \bibinfo
  {author} {\bibfnamefont {J.~M.}\ \bibnamefont {Chow}}, \bibinfo {author}
  {\bibfnamefont {J.}~\bibnamefont {Koch}}, \bibinfo {author} {\bibfnamefont
  {J.~M.}\ \bibnamefont {Gambetta}}, \bibinfo {author} {\bibfnamefont {D.~I.}\
  \bibnamefont {Schuster}}, \bibinfo {author} {\bibfnamefont {L.}~\bibnamefont
  {Frunzio}}, \bibinfo {author} {\bibfnamefont {M.~H.}\ \bibnamefont
  {Devoret}}, \bibinfo {author} {\bibfnamefont {S.~M.}\ \bibnamefont {Girvin}},
  \ and\ \bibinfo {author} {\bibfnamefont {R.~J.}\ \bibnamefont {Schoelkopf}},\
  }\href {\doibase 10.1103/PhysRevLett.101.080502} {\bibfield  {journal}
  {\bibinfo  {journal} {Phys. Rev. Lett.}\ }\textbf {\bibinfo {volume} {101}},\
  \bibinfo {pages} {080502} (\bibinfo {year} {2008})}\BibitemShut {NoStop}%
\bibitem [{\citenamefont {Barends}\ \emph {et~al.}(2013)\citenamefont
  {Barends}, \citenamefont {Kelly}, \citenamefont {Megrant}, \citenamefont
  {Sank}, \citenamefont {Jeffrey}, \citenamefont {Chen}, \citenamefont {Yin},
  \citenamefont {Chiaro}, \citenamefont {Mutus}, \citenamefont {Neill},
  \citenamefont {O'Malley}, \citenamefont {Roushan}, \citenamefont {Wenner},
  \citenamefont {White}, \citenamefont {Cleland},\ and\ \citenamefont
  {Martinis}}]{barends2013}%
  \BibitemOpen
  \bibfield  {author} {\bibinfo {author} {\bibfnamefont {R.}~\bibnamefont
  {Barends}}, \bibinfo {author} {\bibfnamefont {J.}~\bibnamefont {Kelly}},
  \bibinfo {author} {\bibfnamefont {A.}~\bibnamefont {Megrant}}, \bibinfo
  {author} {\bibfnamefont {D.}~\bibnamefont {Sank}}, \bibinfo {author}
  {\bibfnamefont {E.}~\bibnamefont {Jeffrey}}, \bibinfo {author} {\bibfnamefont
  {Y.}~\bibnamefont {Chen}}, \bibinfo {author} {\bibfnamefont {Y.}~\bibnamefont
  {Yin}}, \bibinfo {author} {\bibfnamefont {B.}~\bibnamefont {Chiaro}},
  \bibinfo {author} {\bibfnamefont {J.}~\bibnamefont {Mutus}}, \bibinfo
  {author} {\bibfnamefont {C.}~\bibnamefont {Neill}}, \bibinfo {author}
  {\bibfnamefont {P.}~\bibnamefont {O'Malley}}, \bibinfo {author}
  {\bibfnamefont {P.}~\bibnamefont {Roushan}}, \bibinfo {author} {\bibfnamefont
  {J.}~\bibnamefont {Wenner}}, \bibinfo {author} {\bibfnamefont {T.~C.}\
  \bibnamefont {White}}, \bibinfo {author} {\bibfnamefont {A.~N.}\ \bibnamefont
  {Cleland}}, \ and\ \bibinfo {author} {\bibfnamefont {J.~M.}\ \bibnamefont
  {Martinis}},\ }\href {\doibase 10.1103/PhysRevLett.111.080502} {\bibfield
  {journal} {\bibinfo  {journal} {Phys. Rev. Lett.}\ }\textbf {\bibinfo
  {volume} {111}},\ \bibinfo {pages} {080502} (\bibinfo {year}
  {2013})}\BibitemShut {NoStop}%
\bibitem [{\citenamefont {Schreier}\ \emph {et~al.}(2008)\citenamefont
  {Schreier}, \citenamefont {Houck}, \citenamefont {Koch}, \citenamefont
  {Schuster}, \citenamefont {Johnson}, \citenamefont {Chow}, \citenamefont
  {Gambetta}, \citenamefont {Majer}, \citenamefont {Frunzio}, \citenamefont
  {Devoret}, \citenamefont {Girvin},\ and\ \citenamefont
  {Schoelkopf}}]{schreier}%
  \BibitemOpen
  \bibfield  {author} {\bibinfo {author} {\bibfnamefont {J.~A.}\ \bibnamefont
  {Schreier}}, \bibinfo {author} {\bibfnamefont {A.~A.}\ \bibnamefont {Houck}},
  \bibinfo {author} {\bibfnamefont {J.}~\bibnamefont {Koch}}, \bibinfo {author}
  {\bibfnamefont {D.~I.}\ \bibnamefont {Schuster}}, \bibinfo {author}
  {\bibfnamefont {B.~R.}\ \bibnamefont {Johnson}}, \bibinfo {author}
  {\bibfnamefont {J.~M.}\ \bibnamefont {Chow}}, \bibinfo {author}
  {\bibfnamefont {J.~M.}\ \bibnamefont {Gambetta}}, \bibinfo {author}
  {\bibfnamefont {J.}~\bibnamefont {Majer}}, \bibinfo {author} {\bibfnamefont
  {L.}~\bibnamefont {Frunzio}}, \bibinfo {author} {\bibfnamefont {M.~H.}\
  \bibnamefont {Devoret}}, \bibinfo {author} {\bibfnamefont {S.~M.}\
  \bibnamefont {Girvin}}, \ and\ \bibinfo {author} {\bibfnamefont {R.~J.}\
  \bibnamefont {Schoelkopf}},\ }\href {\doibase 10.1103/PhysRevB.77.180502}
  {\bibfield  {journal} {\bibinfo  {journal} {Phys. Rev. B}\ }\textbf {\bibinfo
  {volume} {77}},\ \bibinfo {pages} {180502} (\bibinfo {year}
  {2008})}\BibitemShut {NoStop}%
\bibitem [{\citenamefont {Sheldon}\ \emph {et~al.}(2016)\citenamefont
  {Sheldon}, \citenamefont {Bishop}, \citenamefont {Magesan}, \citenamefont
  {Filipp}, \citenamefont {Chow},\ and\ \citenamefont {Gambetta}}]{sheldon}%
  \BibitemOpen
  \bibfield  {author} {\bibinfo {author} {\bibfnamefont {S.}~\bibnamefont
  {Sheldon}}, \bibinfo {author} {\bibfnamefont {L.~S.}\ \bibnamefont {Bishop}},
  \bibinfo {author} {\bibfnamefont {E.}~\bibnamefont {Magesan}}, \bibinfo
  {author} {\bibfnamefont {S.}~\bibnamefont {Filipp}}, \bibinfo {author}
  {\bibfnamefont {J.~M.}\ \bibnamefont {Chow}}, \ and\ \bibinfo {author}
  {\bibfnamefont {J.~M.}\ \bibnamefont {Gambetta}},\ }\href {\doibase
  10.1103/PhysRevA.93.012301} {\bibfield  {journal} {\bibinfo  {journal} {Phys.
  Rev. A}\ }\textbf {\bibinfo {volume} {93}},\ \bibinfo {pages} {012301}
  (\bibinfo {year} {2016})}\BibitemShut {NoStop}%
\bibitem [{\citenamefont {Van~Harlingen}\ \emph {et~al.}(2004)\citenamefont
  {Van~Harlingen}, \citenamefont {Robertson}, \citenamefont {Plourde},
  \citenamefont {Reichardt}, \citenamefont {Crane},\ and\ \citenamefont
  {Clarke}}]{harlingen}%
  \BibitemOpen
  \bibfield  {author} {\bibinfo {author} {\bibfnamefont {D.~J.}\ \bibnamefont
  {Van~Harlingen}}, \bibinfo {author} {\bibfnamefont {T.~L.}\ \bibnamefont
  {Robertson}}, \bibinfo {author} {\bibfnamefont {B.~L.~T.}\ \bibnamefont
  {Plourde}}, \bibinfo {author} {\bibfnamefont {P.~A.}\ \bibnamefont
  {Reichardt}}, \bibinfo {author} {\bibfnamefont {T.~A.}\ \bibnamefont
  {Crane}}, \ and\ \bibinfo {author} {\bibfnamefont {J.}~\bibnamefont
  {Clarke}},\ }\href {\doibase 10.1103/PhysRevB.70.064517} {\bibfield
  {journal} {\bibinfo  {journal} {Phys. Rev. B}\ }\textbf {\bibinfo {volume}
  {70}},\ \bibinfo {pages} {064517} (\bibinfo {year} {2004})}\BibitemShut
  {NoStop}%
\bibitem [{\citenamefont {C\'orcoles}\ \emph {et~al.}(2013)\citenamefont
  {C\'orcoles}, \citenamefont {Gambetta}, \citenamefont {Chow}, \citenamefont
  {Smolin}, \citenamefont {Ware}, \citenamefont {Strand}, \citenamefont
  {Plourde},\ and\ \citenamefont {Steffen}}]{corcolesRB}%
  \BibitemOpen
  \bibfield  {author} {\bibinfo {author} {\bibfnamefont {A.~D.}\ \bibnamefont
  {C\'orcoles}}, \bibinfo {author} {\bibfnamefont {J.~M.}\ \bibnamefont
  {Gambetta}}, \bibinfo {author} {\bibfnamefont {J.~M.}\ \bibnamefont {Chow}},
  \bibinfo {author} {\bibfnamefont {J.~A.}\ \bibnamefont {Smolin}}, \bibinfo
  {author} {\bibfnamefont {M.}~\bibnamefont {Ware}}, \bibinfo {author}
  {\bibfnamefont {J.}~\bibnamefont {Strand}}, \bibinfo {author} {\bibfnamefont
  {B.~L.~T.}\ \bibnamefont {Plourde}}, \ and\ \bibinfo {author} {\bibfnamefont
  {M.}~\bibnamefont {Steffen}},\ }\href {\doibase 10.1103/PhysRevA.87.030301}
  {\bibfield  {journal} {\bibinfo  {journal} {Phys. Rev. A}\ }\textbf {\bibinfo
  {volume} {87}},\ \bibinfo {pages} {030301} (\bibinfo {year}
  {2013})}\BibitemShut {NoStop}%
\bibitem [{\citenamefont {Paraoanu}(2006)}]{paraoanu}%
  \BibitemOpen
  \bibfield  {author} {\bibinfo {author} {\bibfnamefont {G.~S.}\ \bibnamefont
  {Paraoanu}},\ }\href {\doibase 10.1103/PhysRevB.74.140504} {\bibfield
  {journal} {\bibinfo  {journal} {Phys. Rev. B}\ }\textbf {\bibinfo {volume}
  {74}},\ \bibinfo {pages} {140504} (\bibinfo {year} {2006})}\BibitemShut
  {NoStop}%
\bibitem [{\citenamefont {Gambetta}\ \emph {et~al.}(2012)\citenamefont
  {Gambetta}, \citenamefont {C\'orcoles}, \citenamefont {Merkel}, \citenamefont
  {Johnson}, \citenamefont {Smolin}, \citenamefont {Chow}, \citenamefont
  {Ryan}, \citenamefont {Rigetti}, \citenamefont {Poletto}, \citenamefont
  {Ohki}, \citenamefont {Ketchen},\ and\ \citenamefont {Steffen}}]{gambetta}%
  \BibitemOpen
  \bibfield  {author} {\bibinfo {author} {\bibfnamefont {J.~M.}\ \bibnamefont
  {Gambetta}}, \bibinfo {author} {\bibfnamefont {A.~D.}\ \bibnamefont
  {C\'orcoles}}, \bibinfo {author} {\bibfnamefont {S.~T.}\ \bibnamefont
  {Merkel}}, \bibinfo {author} {\bibfnamefont {B.~R.}\ \bibnamefont {Johnson}},
  \bibinfo {author} {\bibfnamefont {J.~A.}\ \bibnamefont {Smolin}}, \bibinfo
  {author} {\bibfnamefont {J.~M.}\ \bibnamefont {Chow}}, \bibinfo {author}
  {\bibfnamefont {C.~A.}\ \bibnamefont {Ryan}}, \bibinfo {author}
  {\bibfnamefont {C.}~\bibnamefont {Rigetti}}, \bibinfo {author} {\bibfnamefont
  {S.}~\bibnamefont {Poletto}}, \bibinfo {author} {\bibfnamefont {T.~A.}\
  \bibnamefont {Ohki}}, \bibinfo {author} {\bibfnamefont {M.~B.}\ \bibnamefont
  {Ketchen}}, \ and\ \bibinfo {author} {\bibfnamefont {M.}~\bibnamefont
  {Steffen}},\ }\href {\doibase 10.1103/PhysRevLett.109.240504} {\bibfield
  {journal} {\bibinfo  {journal} {Phys. Rev. Lett.}\ }\textbf {\bibinfo
  {volume} {109}},\ \bibinfo {pages} {240504} (\bibinfo {year}
  {2012})}\BibitemShut {NoStop}%
\bibitem [{\citenamefont {Magesan}\ \emph
  {et~al.}(2012{\natexlab{a}})\citenamefont {Magesan}, \citenamefont
  {Gambetta}, \citenamefont {Johnson}, \citenamefont {Ryan}, \citenamefont
  {Chow}, \citenamefont {Merkel}, \citenamefont {da~Silva}, \citenamefont
  {Keefe}, \citenamefont {Rothwell}, \citenamefont {Ohki}, \citenamefont
  {Ketchen},\ and\ \citenamefont {Steffen}}]{magesanIRB}%
  \BibitemOpen
  \bibfield  {author} {\bibinfo {author} {\bibfnamefont {E.}~\bibnamefont
  {Magesan}}, \bibinfo {author} {\bibfnamefont {J.~M.}\ \bibnamefont
  {Gambetta}}, \bibinfo {author} {\bibfnamefont {B.~R.}\ \bibnamefont
  {Johnson}}, \bibinfo {author} {\bibfnamefont {C.~A.}\ \bibnamefont {Ryan}},
  \bibinfo {author} {\bibfnamefont {J.~M.}\ \bibnamefont {Chow}}, \bibinfo
  {author} {\bibfnamefont {S.~T.}\ \bibnamefont {Merkel}}, \bibinfo {author}
  {\bibfnamefont {M.~P.}\ \bibnamefont {da~Silva}}, \bibinfo {author}
  {\bibfnamefont {G.~A.}\ \bibnamefont {Keefe}}, \bibinfo {author}
  {\bibfnamefont {M.~B.}\ \bibnamefont {Rothwell}}, \bibinfo {author}
  {\bibfnamefont {T.~A.}\ \bibnamefont {Ohki}}, \bibinfo {author}
  {\bibfnamefont {M.~B.}\ \bibnamefont {Ketchen}}, \ and\ \bibinfo {author}
  {\bibfnamefont {M.}~\bibnamefont {Steffen}},\ }\href {\doibase
  10.1103/PhysRevLett.109.080505} {\bibfield  {journal} {\bibinfo  {journal}
  {Phys. Rev. Lett.}\ }\textbf {\bibinfo {volume} {109}},\ \bibinfo {pages}
  {080505} (\bibinfo {year} {2012}{\natexlab{a}})}\BibitemShut {NoStop}%
\bibitem [{\citenamefont {Gaebler}\ \emph {et~al.}(2012)\citenamefont
  {Gaebler}, \citenamefont {Meier}, \citenamefont {Tan}, \citenamefont
  {Bowler}, \citenamefont {Lin}, \citenamefont {Hanneke}, \citenamefont {Jost},
  \citenamefont {Home}, \citenamefont {Knill}, \citenamefont {Leibfried},\ and\
  \citenamefont {Wineland}}]{gaebler}%
  \BibitemOpen
  \bibfield  {author} {\bibinfo {author} {\bibfnamefont {J.~P.}\ \bibnamefont
  {Gaebler}}, \bibinfo {author} {\bibfnamefont {A.~M.}\ \bibnamefont {Meier}},
  \bibinfo {author} {\bibfnamefont {T.~R.}\ \bibnamefont {Tan}}, \bibinfo
  {author} {\bibfnamefont {R.}~\bibnamefont {Bowler}}, \bibinfo {author}
  {\bibfnamefont {Y.}~\bibnamefont {Lin}}, \bibinfo {author} {\bibfnamefont
  {D.}~\bibnamefont {Hanneke}}, \bibinfo {author} {\bibfnamefont {J.~D.}\
  \bibnamefont {Jost}}, \bibinfo {author} {\bibfnamefont {J.~P.}\ \bibnamefont
  {Home}}, \bibinfo {author} {\bibfnamefont {E.}~\bibnamefont {Knill}},
  \bibinfo {author} {\bibfnamefont {D.}~\bibnamefont {Leibfried}}, \ and\
  \bibinfo {author} {\bibfnamefont {D.~J.}\ \bibnamefont {Wineland}},\ }\href
  {\doibase 10.1103/PhysRevLett.108.260503} {\bibfield  {journal} {\bibinfo
  {journal} {Phys. Rev. Lett.}\ }\textbf {\bibinfo {volume} {108}},\ \bibinfo
  {pages} {260503} (\bibinfo {year} {2012})}\BibitemShut {NoStop}%
\bibitem [{\citenamefont {Magesan}\ \emph
  {et~al.}(2012{\natexlab{b}})\citenamefont {Magesan}, \citenamefont
  {Gambetta},\ and\ \citenamefont {Emerson}}]{magesan12}%
  \BibitemOpen
  \bibfield  {author} {\bibinfo {author} {\bibfnamefont {E.}~\bibnamefont
  {Magesan}}, \bibinfo {author} {\bibfnamefont {J.~M.}\ \bibnamefont
  {Gambetta}}, \ and\ \bibinfo {author} {\bibfnamefont {J.}~\bibnamefont
  {Emerson}},\ }\href {\doibase 10.1103/PhysRevA.85.042311} {\bibfield
  {journal} {\bibinfo  {journal} {Phys. Rev. A}\ }\textbf {\bibinfo {volume}
  {85}},\ \bibinfo {pages} {042311} (\bibinfo {year}
  {2012}{\natexlab{b}})}\BibitemShut {NoStop}%
\bibitem [{\citenamefont {{Ball}}\ \emph {et~al.}(2016)\citenamefont {{Ball}},
  \citenamefont {{Oliver}},\ and\ \citenamefont {{Biercuk}}}]{ball}%
  \BibitemOpen
  \bibfield  {author} {\bibinfo {author} {\bibfnamefont {H.}~\bibnamefont
  {{Ball}}}, \bibinfo {author} {\bibfnamefont {W.~D.}\ \bibnamefont
  {{Oliver}}}, \ and\ \bibinfo {author} {\bibfnamefont {M.~J.}\ \bibnamefont
  {{Biercuk}}},\ }\href@noop {} {\bibfield  {journal} {\bibinfo  {journal}
  {ArXiv e-prints}\ } (\bibinfo {year} {2016})},\ \Eprint
  {http://arxiv.org/abs/1602.04551} {arXiv:1602.04551 [quant-ph]} \BibitemShut
  {NoStop}%
\end{thebibliography}%

\end{document}